# Chemical-Vapor-Deposited Graphene as a Thermally Conducting Coating

Mauro Tortello*,1, Iwona Pasternak2, Klaudia Zeranska-Chudek2, Wlodek Strupinski2, Renato S. Gonnelli1,

Alberto Fina1

1. Dipartimento di Scienza Applicata e Tecnologia, Politecnico di Torino, 10129 Torino, Italy

2. Faculty of Physics, Warsaw University of Technology, Koszykowa 75, 00-662 Warsaw Poland

We performed Scanning Thermal Microscopy measurements on single layers of chemical-vapor-deposited (CVD) graphene supported by different substrates, namely $SiO_2$, $Al_2O_3$ and PET using a double-scan technique to remove the contribution to the heat flux through the air and the cantilever. Then, by adopting a simple lumped-elements model, we developed a new method that allows determining, through a multi-step numerical analysis, the equivalent thermal properties of thermally conductive coatings of nanometric thickness. In this specific case we found that our CVD graphene is "thermally equivalent", for heat injection perpendicular to the graphene planes, to a coating material of conductivity $k_{eff} = 2.5 \pm 0.3 \frac{W}{mK}$ and thickness $t_{eff} = 3.5 \pm 0.3\ nm$ *in perfect contact* with the substrate. For the $SiO_2$ substrate, we also measured stacks made of 2- and 4- CVD monolayers and we found that the effective thermal conductivity increases with increasing number of layers and, with a technologically achievable number of layers, is expected to be comparable to that of one order of magnitude-thicker metallic thin films. This study provides a powerful method for characterizing the thermal properties of graphene in view of several thermal management applications.

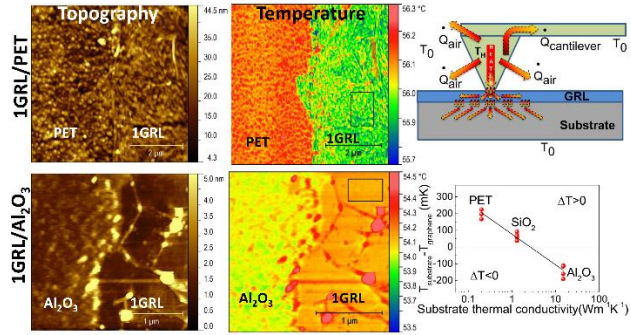

**INTRODUCTION**

It is known that the remarkable electrical [1-3] and thermal [4-7] properties of graphene can change considerably depending on its quality and on the specific system in which graphene is employed. Indeed, the number of layers [8-10], amount of defects [11-15], coupling to the substrate [16,17], production method [18], presence of graphene-substrate adsorbate layer or water adlayers [19,20,21], etc., can give rise to different electrical and thermal properties and/or performances. For example, the exceptionally high thermal conductivity of suspended, mechanically exfoliated graphene decreases by one order of magnitude when it is supported by SiO$_2$, due to the coupling of the flexural ZA vibrational modes to the substrate [22]. Moreover, the thermal conductivity of single layer graphene has also been shown to have a 30% to 50% reduction in an epoxy matrix [23]. Therefore, it is very important to evaluate and investigate the properties of graphene or graphene-related materials (but this consideration holds for all 2D materials) in the specific system in which they have to be employed.

In the perspective of utilizing graphene in future (possibly flexible) electronics, it is very important to consider the thermal conductivity, heat generation and dissipation of supported (rather than suspended) graphene and its interaction with different substrates, since the performance of electronic devices considerably depends on the temperature [24]. For applications like thermally conductive nanocomposites [25,26], thermal interface materials [27,28,29], thermally conductive coatings for plastic materials [30] and innovative heat spreaders [27,28,31], the interaction between graphene and oxides (like $SiO_2$), metals or polymers can be crucial. Furthermore, the investigation of the thermal conductivity properties of CVD graphene is much more relevant to applications compared to exfoliated graphene, as large-scale CVD processes are currently available and exploited for thin film industrial applications [32].

Scanning Thermal Microscopy (SThM) [33,34] is a powerful technique for investigating the thermal properties at the nanoscale. Despite this technique hardly provides a quantitative determination of the thermal conductivity of the sample [34,35], SThM has an unmatched spatial resolution (a few tens of nanometers or less), which cannot be achieved by other popular methods such as the Raman optothermal technique [10] or by electrical methods [19].

By performing SThM measurements, Pumarol et al. [36] showed that the heat transport in suspended exfoliated graphene is higher than for the supported one and that the thermal conductance per single layer in a 3-layer graphene is about 68% of that of supported single layer graphene. Menges et al. [37] measured single and multilayer graphene supported by $SiO_2$ or crystalline SiC and claimed a sub-10 nm lateral resolution with a thickness sensitivity to the single atomic layer. Furthermore, they observed a decrease of the thermal resistance with increasing number of layers for $SiO_2$-supported, mechanically exfoliated graphene. A 30 nm spatial resolution was reported by Tovee et al. on few-layer graphene by using carbon nanotube tipped thermal probes [38]. K. Yoon et al. [39] quantitatively determined the thermal conductivity of suspended graphene by using the so called null-point SThM that employs a thermocouple as the thermal probe. In this work and in others [40], however, the authors do not usually report thermal maps but only line scans. Tortello et al. reported on the thermal properties of pristine and annealed reduced graphite oxide flakes [35]: a correlation between the reduction of structure defectiveness consequent to annealing and improved thermal properties was demonstrated by SThM measurements on the single flakes.

To the best of our knowledge, no SThM studies of graphene grown by chemical vapor deposition (CVD) were previously reported, despite this is currently the best candidate for large-scale production of graphene-based devices, since mechanical exfoliation, that gives the best samples in terms of quality, is certainly not viable in this regard.

Here we show SThM results on CVD graphene (1GRL) supported by different substrates i.e. $SiO_2$, Polyethylene terephthalate (PET) and $Al_2O_3$. For the $SiO_2$ substrate we also measured samples with 2 (2GRL) and 4 (4GRL) CVD graphene layers stacked one on top of the other (random stacking).

**EXPERIMENTAL SECTION**

**Scheme 1. a: Temperature vs time diagram of the CVD graphene growth process as described in [41]. b: Sequence of the steps for the marker-frame method used for the transfer of CVD graphene on the different substrates [44].**

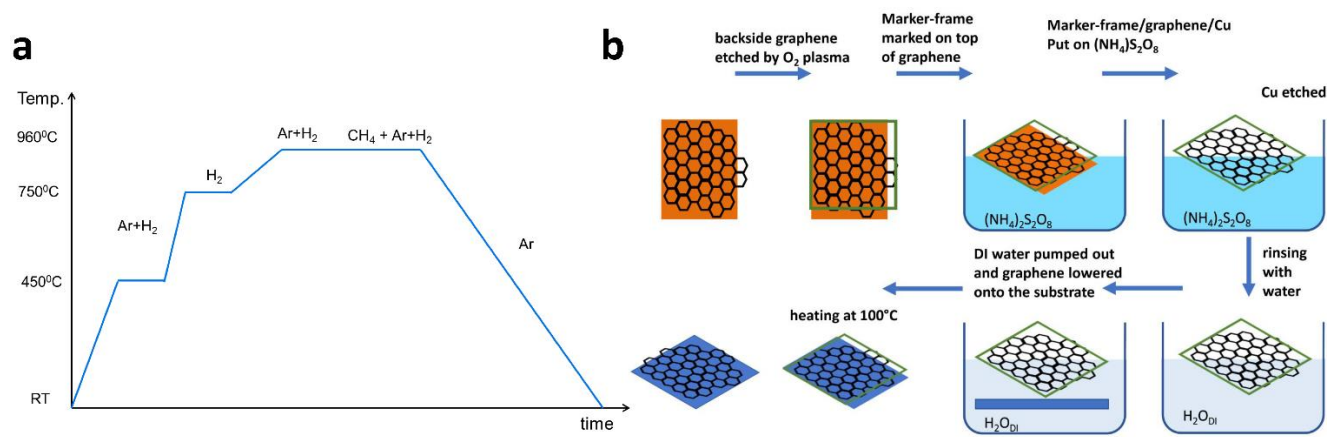

The graphene films were grown by chemical vapor deposition (CVD) on top of 25 μm-thick copper substrates, as described in Ciuk et al. [41]. The temperature vs time diagram followed for the growth process is reported in Scheme **1a**. Bi or tri-layers of graphene on the original graphene film are usually observed as 1-2 μm hexagons or dendrites scattered on the surface. These layers are presumed to grow underneath the first layer at the same copper active site (impurities) as the first layer [42,43]. These areas can be seen as dark spots in SEM images or as bright spots in optical images. We avoided these regions during SThM measurements, as it will be shown later. The graphene films were then transferred to different substrates by using a special marker-frame method (Scheme **1b**) that does not make use of polymers like PMMA or PDMS, thus avoiding leaving polymer residues. [44]. Moreover, this method allows transferring the graphene films on almost any substrate, since there is no need of using dissolving agents, like e.g. acetone, normally employed for removing polymers. Three different substrates were adopted, PET, 285nm silicon dioxide grown on silicon (SiO$_2$/Si) and alumina (Al$_2$O$_3$). The SiO$_2$ substrate was a dry thermal oxide while the Al$_2$O$_3$ one was monocrystalline Epi-ready sapphire. On each of them, we transferred 1 graphene layer (1GRL). In the case of SiO$_2$/Si substrate, we also prepared samples with two (2GRL) and four (4GRL) layers. The different substrates were chosen to span in thermal conductivity by two orders of magnitude ($k_{PET} = 0.2 \frac{W}{mK}$, $k_{SiO2} = 1.4 \frac{W}{mK}$ and $k_{Al2O3} = 15 \frac{W}{mK}$). The samples were characterized by Raman spectroscopy using a Renishaw inVia system and a wavelength of 514 nm. It also worth pointing out here that, unless the samples are prepared in dry conditions (which is not the case here), it has been shown that there is a ubiquitous graphene-substrate adsorbate layer [19,20,21] that will tend to make the interface properties similar among different substrates. For this reason, we will later assume that, to a first approximation, the thermal contact resistance between the graphene and the substrate is the same for all the substrates.

Scanning Thermal Microscopy (SThM) measurements were performed on an Innova© atomic force microscope (AFM) from Bruker, equipped with a VITA module for the thermal measurements. For the SThM measurements we adopted state-of-the-art resistive probes (Bruker VITA HE-GLA) in which a thin Pd film is deposited near the silicon nitride probe apex. The thin film acts at the same time as the heater and the temperature sensor and is part of a Wheatstone bridge. Before the measurements, the resistance of the probe is first measured (by means of an Agilent 34420A nano-voltmeter) at a low current value, i.e. 100 μA, to avoid Joule heating and subsequently at a higher value (1 mA) at which the probe is heated. This is necessary to obtain the value of the resistance because it can slightly change over time (days) of repeated measurements. Then, the measured value is compared to that obtained by using the standard Wheatstone bridge formula that requires, as the input, the bridge voltage provided by the instrument software. This operation is necessary to check that the formula is providing the correct resistance value, since these values in the SThM measurements will be obtained through the mentioned procedure. The heating effect due to the laser is also considered by repeating the procedure first with the laser off and then with the laser turned on. The thermal scans are then performed by applying a current of about 1.3-1.4 mA, since higher values are likely to alter the resistance or even damage the probe. Then, after a thermal map has been acquired, the bridge voltage is converted to a resistance value and the resistance is converted to temperature by using the temperature coefficient of the probes, that we measured to be $8.92 \times 10^{-4} K^{-1}$, similar to that reported for Palladium [48] but lower than the one measured on the older generation of probes, made of silicon dioxide [49]. The temperature increase due to the laser is normally of about 0.8-1.2 K. The probe is formed by two NiCr "legs" resistors deposited on the cantilever and by the heater part formed by the Pd resistor at the tip apex. Indeed, since the temperature coefficient of Pd is one order of magnitude higher than that of NiCr while their electrical resistances are comparable (around 100 Ω each), we assumed that the temperature coefficient of the resistive part close to the apex is that of the whole probe. This is confirmed by the fact that the total temperature coefficient that we determined differs by less than 5% from that of pure Pd. Thus, we can, to a good extent, consider that most of the temperature variation is occurring at the tip apex that is also hotter than the rest of the probe. Therefore, in the following we will consider that the resistive sensor is localized only at the tip apex.

The SThM tips that we employed are state-of-the-art microfabricated probes. We think it is not yet technologically possible to obtain this kind of probes with a higher aspect ratio together with the required fabrication repeatability (especially considering the presence of the Pd resistive film deposited on the tip apex). To the best of our knowledge, a better resolution has been claimed for the silicon probes [37], but the heater is farther from the sample and our AFM has been optimized for the Pd probes that we adopted. Another possibility to enhance the resolution, could be to attach a carbon nanotube to the probe, as it has been done by Tovee et al. [38]. This would be interesting but rather beyond the scope of this paper where we are more interested in a reliable method for determining the thermal properties of 2D materials for heat injection along the cross-plane direction.

In the SThM measurements, a lower temperature of the sensor means that a higher heat flux is transferred from the probe to the sample with respect to a region where the temperature is higher. The average temperature in a certain region is obtained by applying a mask and by averaging the temperature of each pixel contained in the mask. The temperature difference between the substrate and the graphene is $T_{sub} - T_{GR} =$ $\Delta T$. The temperature uncertainty on each mask, $\delta T$ is determined by the standard deviation and the final uncertainty is determined by the propagation of the error on each temperature, i.e. $\delta \Delta T = \sqrt{\delta T_{sub}^2 + \delta T_{GR}^2}$. From the instrumental point of view, the minimum resolution in the bridge voltage corresponds to a temperature variation of about 1 mK, which is however not corresponding to the actual achievable precision due to various sources of environmental noise (thermal, electrical etc.). Indeed, the uncertainty on the temperature determination on different areas of the sample will be of the order of some tens of mK. We also point out here that results similar to those obtained with the masking procedure can be obtained by applying a thresholding method in order to single out the flat areas of the sample in the same temperature range. Finally, by knowing the ambient temperature, $T_0$ and the applied power, P (determined by the Joule-heating formula, $P=R_H I^2$) the maps of the total thermal resistance of the systems can also be obtained.

The SThM measurements are performed in the contact mode and the topography and other typical signals of this mode, like the lateral force, can be recorded while at the same time acquiring the thermal maps. The lateral force was found to be very powerful for clearly distinguishing between the graphene and the substrate regions.

## RESULTS AND DISCUSSION

i. 1, 2 and 4 layers supported by SiO$_2$/Si

Figure **1** shows Raman spectra of graphene layers transferred onto SiO$_2$/Si substrates. Raman spectra indicate two prominent and characteristic G and 2D peaks which are the features confirming the presence of graphene. The disorder-related weak D peaks connected with defects are also present. For the spectrum marked as "1GRL", the observed narrow (with the full width at half maximum (FWHM) of 35 cm$^{-1}$) and symmetric Lorentzian lineshape of the 2D peak is a feature confirming the presence of predominantly single layer graphene [45]. For the "2GRL" and "4GRL" we note a broadening of the 2D band and a slight shift of its position. These observations confirm that the shape and frequency of the 2D band are sensitive to the number of graphene layers. Indeed, in the case of exfoliated graphene (with defined stacking order) they can be used to determine the exact number of layers [46]. However, regarding our experiments where the graphene layers were added one by one, the created multilayer stack is in random alignment configurations [47] and, therefore, it is not possible to determine the number of graphene layers by analyzing the 2D peak.

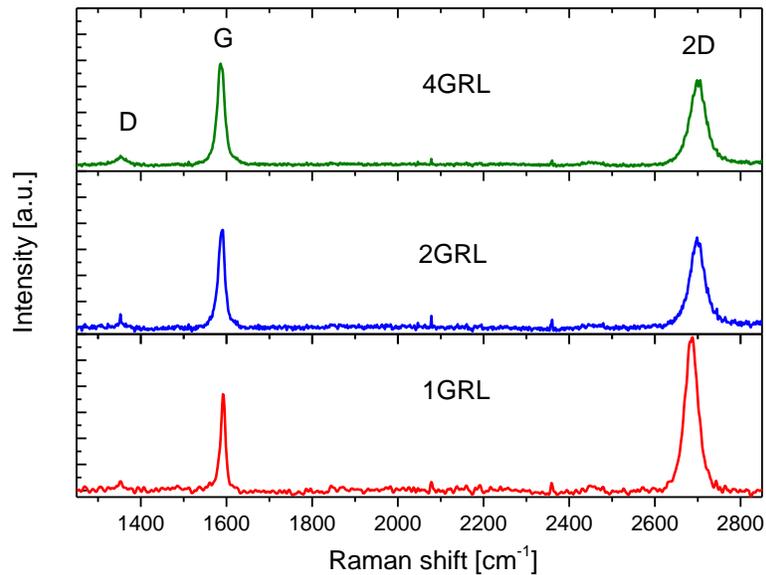

Figure **1**. Raman spectra of 1GRL, 2GRL and 4GRL on SiO$_2$/Si substrates.

Figure **2**a shows the topography map of 1GRL supported by a SiO$_2$/Si substrate. The graphene is covering the lower-left half of the image, but it is hardly distinguishable from the substrate also owing to the negligible thickness of graphene as compared with the height of some impurities saturating the scale. The presence of several wrinkles in that region, however, approximately indicates where the monolayer is located. The origin of the wrinkles is twofold. First, graphene was transferred from a copper foil. It is well known that due to the mismatch of the thermal expansion between graphene and copper, the graphene ripples [50]. Second, wrinkles might come from the method of graphene transfer. In the marker-frame method, the graphene almost freely floats on a water surface, and such fluctuations can foster graphene wrinkling. Additionally, the standard procedure of graphene transfer includes annealing at 300-400°C to flatten the ripples. Since we transferred graphene also on PET foil which is not resistant to those temperatures, in the case of our experiments we decided to skip this step and we kept the same conditions for all substrates. The lateral force signal (panel b), on the other hand, clearly and unambiguously shows the presence of the graphene layer, since the friction between the probe and the sample is very different for the graphene or the substrate. Panel c represents the corresponding thermal map. It is possible to see that the temperature of the sensor is lower when the probe is in contact with the graphene layer than when it is on the bare substrate. The temperature on the graphene is determined by the average temperature of the masked unwrinkled region (rectangle in panel c), while the temperature on the substrate is determined by a similar mask placed on the substrate (not shown). The temperature difference between the substrate and the graphene is $T_{sub} - T_{GR} = \Delta T = 92 \pm 44$ mK. This temperature difference indicates that a greater heat flux is present when the probe is on the graphene than when it is on the substrate. It is also worth noticing here that the temperature has to be determined on the flat areas of the samples, in order to avoid "topological artifacts" [51]. Indeed, when the probe is, for instance, on the top of a significantly higher and steep region (like the impurities that are shown in red color in the lower-right part of panel c), a lower heat flux is transferred to the sample (via conduction through the air) because the distance from the sample has increased with respect to a flat area and the sensor temperature increases. On the other hand, when the probe is inside a concave structure, air-mediated heat transfer contribution becomes higher, increasing the total heat dissipation and consequently decreasing the sensor temperature. In this regard, the small, higher temperature spot at the center of the mask of panel c was excluded from the average temperature calculation. By looking at the thermal maps, one might also wonder how the thermal

conductivity behaves at defects and, especially, at line defects and whether it is possible to resolve its behavior. In this regard, we expect of course a decrease of the thermal conduction properties at defects locations due to increased phonon scattering, but one of the experimental limitations will be the spatial resolution of the tip. The resolution of these probes is around 20-30 nm, thus not enough, in principle, to resolve a line defect, which occurs on a much smaller distance. It might nevertheless be possible that, while scanning over a line defect, a small increase of the temperature is detected. However, this experiment should necessarily be performed on graphene samples deposited on atomically flat substrates, e.g. h-BN. Indeed, for detecting a change in thermal conduction over such a small length scale we should get rid of all possible topological artefacts that might give an apparent temperature variation. Finally, we point out that the scanning direction should also be perpendicular to the line defect because the noise along the scanning direction is lower than between adjacent scan lines. This might help to observe a temperature increase along each scan line in the point where the tip passes over the defects.

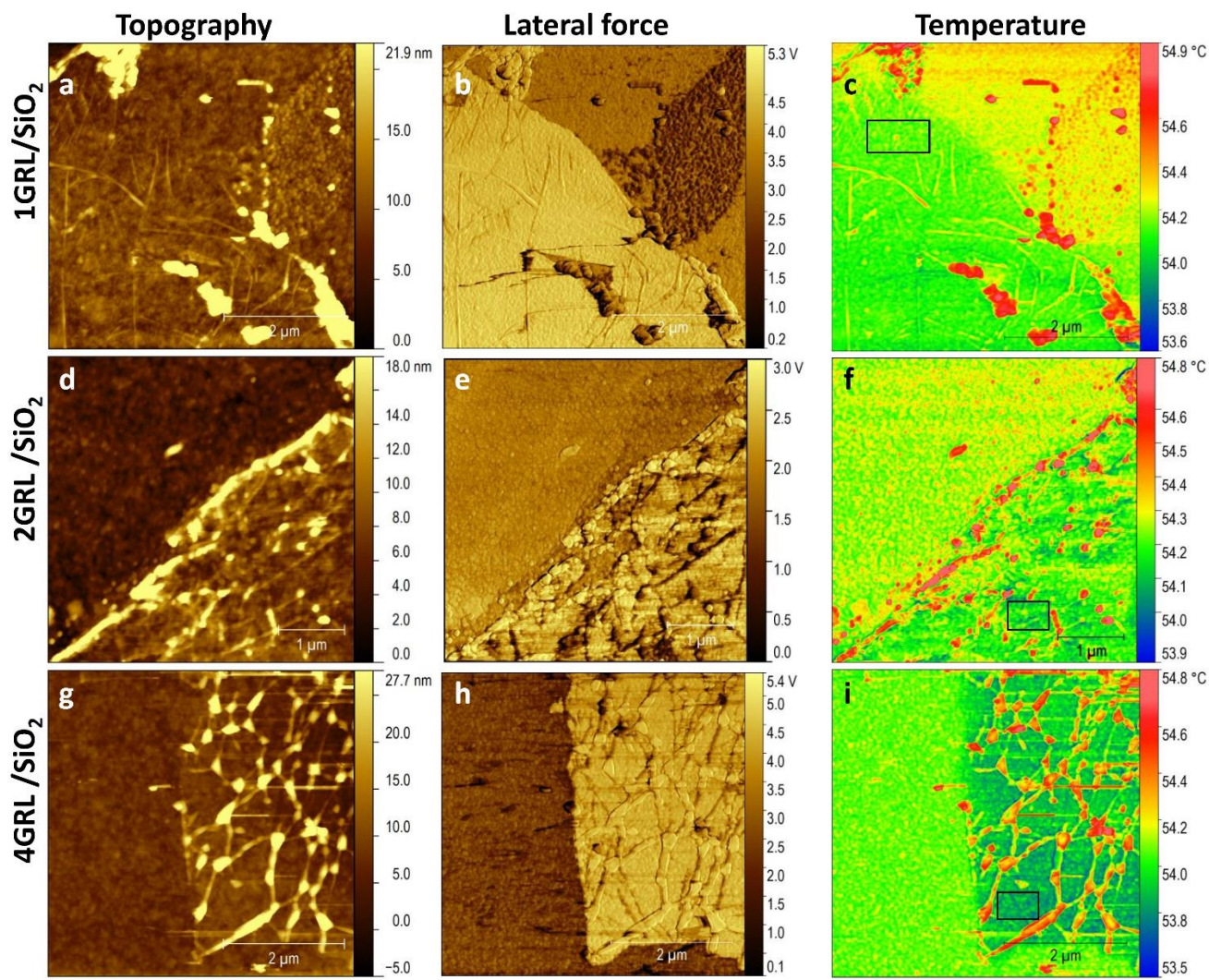

Figure 2. a, b and c: Topography, lateral force and SThM maps of 1GRL supported by SiO$_2$/Si substrate, respectively. d, e and f: The same as for a, b and c but for 2GRL. g, h and i: The same as for a, b and c but for 4GRL.

Panel d reports the topography map of 2GRL on SiO$_2$/Si. In this case it is easier to identify the graphene sample, mainly because of the presence of some impurities, especially located at its edge, related to residues of chemicals used in the graphene transfer process. Moreover, as for the case of the 1GRL sample, we can also notice here some wrinkles on its surface. A clear contrast is observed in the lateral force map (panel e) also showing that the surface of the sample is in this case less homogeneous and presents a few more irregularities compared to the 1GRL sample. The impurities are also very well highlighted in the thermal map (panel f) due to the above-mentioned topological effects. However, several flat regions are present where the temperature can be reliably determined, as in the area indicated by the rectangular mask. By calculating the average temperature on a similar area on the substrate, we obtained for this sample $\Delta T = 111 \pm 69$ mK, which is slightly higher than that observed for the 1GRL sample.

Panel g shows the results of 4GRL on $SiO_2$/Si sample. The sample is characterized by several flat, tile-like areas, surrounded by wrinkles, rather noticeable. This morphology is even more clearly indicated by the lateral force image (panel h). These structures are rather pronounced and look very similar to those reported by Kretinin et al. [52] and might be related to inclusions of organic residues. However, since the SThM probe is injecting the heat and measuring the temperature locally, their contribution to the thermal conduction is confined to the defective regions and their effect may easily be excluded by the proper selection of the analysis areas. Panel k reports the temperature map where we can see, at the same time, a clear temperature contrast between the flat areas and the substrate and the presence, as expected, of high-temperature regions in correspondence of the folds. The temperature contrast obtained in this case is $\Delta T = 221 \pm 65$ mK, clearly higher than for the 1GRL and 2GRL case, indicating that a higher heat flux is dissipated from the tip through the sample. Finally, it is worth noticing that all the thermal maps shown here do not present any lower-temperature area with a size of 1 or 2 μm, that could be compatible with the possible presence of bi- or tri-layer regions formed during the growth process.

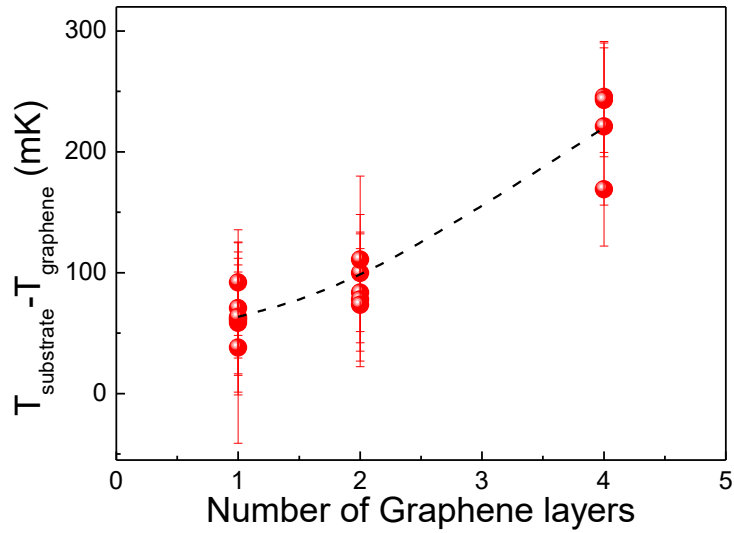

Figure 3. Summary of the temperature difference $T_{sub} - T_{GR} = \Delta T$ between the sensor temperature with the probe on the substrate and on graphene, as a function of the number of graphene layers. The dashed line is only a guide to the eye.

Figure 3 reports, for the three cases, a summary of the temperature contrasts obtained by scanning on different areas of the samples. Even though the $\Delta T$ values are affected by a significant experimental error band, a clear trend is visible where the temperature contrast increases with the number of layers. The average values are $\overline{\Delta T} = 64 \pm 27\ mK$, $\overline{\Delta T} = 89 \pm 19\ mK$ and $\overline{\Delta T} = 220 \pm 39\ mK$ for the 1GRL, 2GRL and 4GRL samples, respectively.

To analyze the data and discuss the results, we adopt the simplest lumped-elements circuit model for the heat conduction in this system, in a similar way as reported in other works [35,34,53] and as shown in Figure **4**.

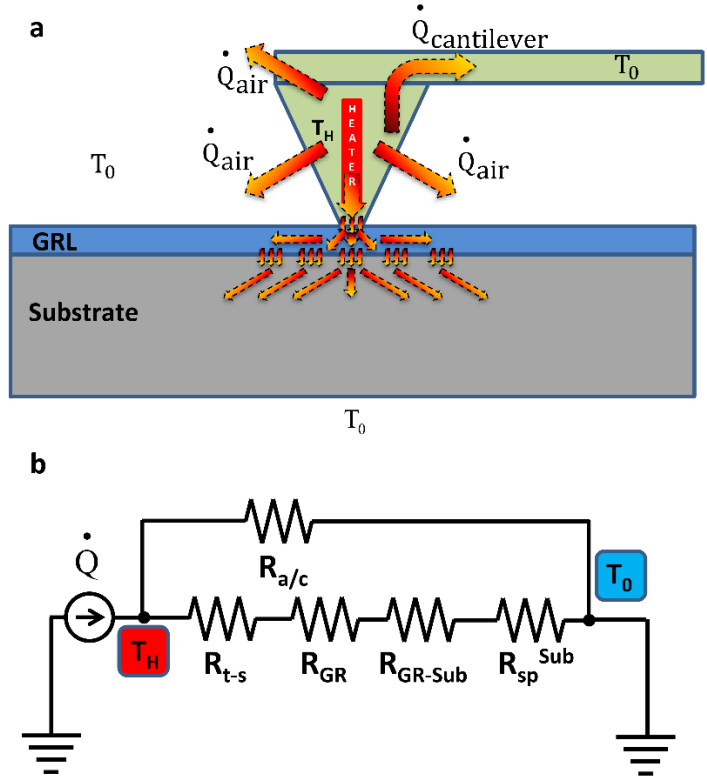

Figure 4. a: Sketch of the SThM probe in contact with a supported graphene sample. The arrows represent the different heat conduction channels between the heater at temperature $T_H$ and the ambient temperature at $T_0$. b: Equivalent lumped-elements circuit model for the heat conduction paths of the system sketched in a.

ii. *Lumped-elements model*

The thermal resistance is defined as $R = \frac{T_H - T_0}{\dot{Q}}$ where $T_H$ is the temperature of the hot region (i.e. the heater), $T_0$ is the ambient temperature and $\dot{Q}$ is the heat flux between them. When the probe is on the graphene, the total thermal conductance can be written as $\frac{1}{R_{tot}} = \frac{1}{R_{a/c}} + \frac{1}{R_{t-s} + R_{GR} + R_{GR-Sub} + R_{sp}^{Sub}}$ where $R_{a/c}$ describes the heat dissipation from both the heater to the air and from the heater through the cantilever, $R_{t-s}$ is the contact resistance between the tip and the sample, $R_{GR}$ is the resistance of the graphene sample, $R_{GR-Sub}$ is the thermal boundary resistance between the substrate and graphene and $R_{sp}^{Sub}$ represents the spreading resistance through the substrate. The different heat conduction paths are represented in the schematic of the probe shown in Figure **4** a. In the thermal maps reported in Figures **2** c, f and i the only difference is the number of graphene layers (1, 2 and 4, respectively). Therefore, the only quantity that changes from one case to the other is $R_{GR}$. Since the temperature contrast between the substrate and the sample increases with increasing number of layers, as reported in Figure **3**, $R_{GR}$ decreases with increasing number of layers when passing from 1 to 4 layers. This result agrees with what has been reported for SThM measurements on exfoliated graphene [37].

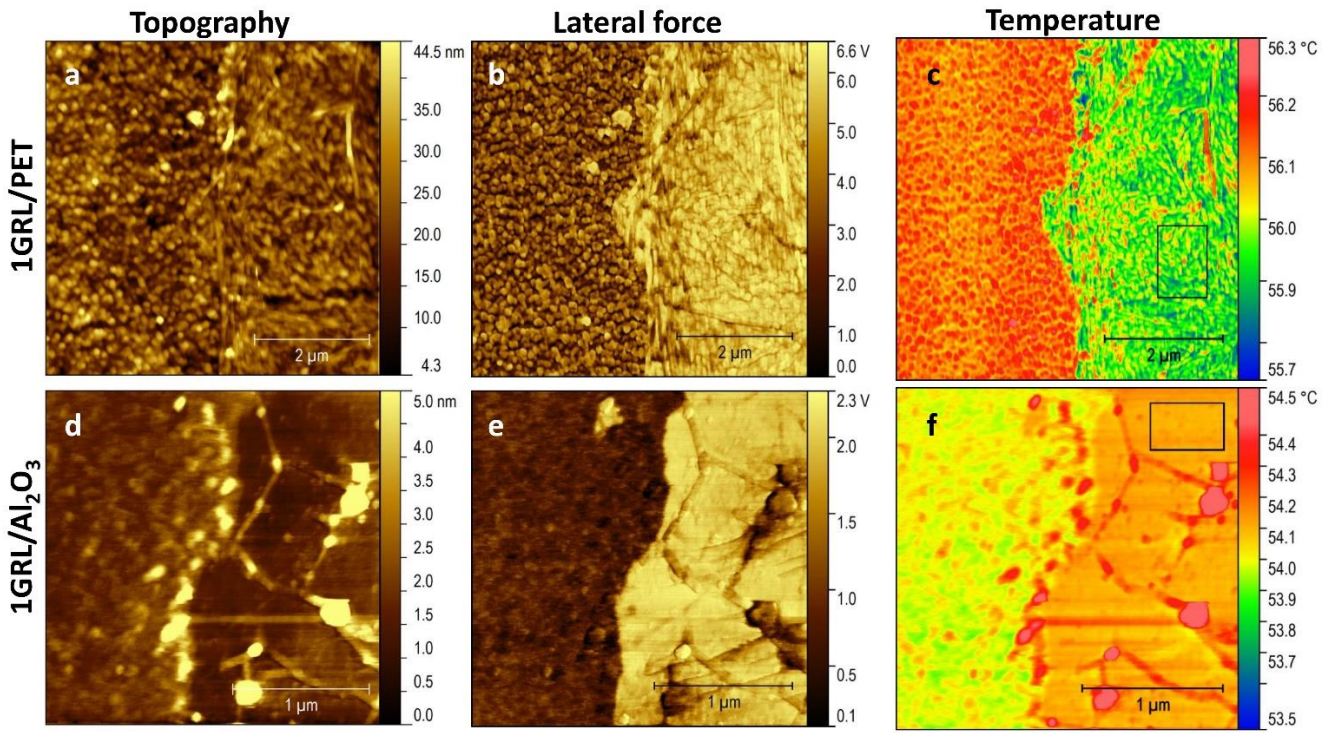

Figure **5.** a, b and c: Topography, lateral force and SThM maps of 1GRL supported by PET substrate, respectively. d, e and f: The same as for a, b and c but for an Al$_2$O$_3$ substrate.

Figure **5** a reports the topography image for the 1GRL supported by PET. Again, it is not easy to identify the graphene layer, but the presence of some wrinkles suggests that the right part of the area is covered by the 2D monolayer. Indeed, this is confirmed unambiguously by the lateral force map (panel b) that also in this case shows an evident difference between the graphene and the substrate. Panel c shows the thermal map where a clear temperature contrast between the graphene and the PET region can be observed, the second featuring a higher temperature. The temperature difference we obtained in this case is $T_{sub} - T_{GR} = \Delta T = 167 \pm 64$ mK, which is significantly higher than for the monolayer on the SiO$_2$/Si substrate. An enhancement of the temperature contrast when passing from the SiO$_2$/Si to the PET substrate was also observed in the case of SThM measurements of graphite nanoplates with thickness in the 4-15 nm range [35]. Panel d shows the topography of a graphene layer supported by Al$_2$O$_3$. The graphene is located at the right-hand side of the image, as confirmed by the lateral force map of panel e. As for the thermal map, a clear temperature contrast is observed also in this case but with a significant difference: unlike the previous cases, the sensor temperature is now *higher* when the probe is on the graphene than when it is on the substrate, with $\Delta T = -110 \pm 32$ mK. This is clearly related to the thermal conductivity of the substrate, which for alumina is approximately one order of magnitude higher than for SiO$_2$. The change in sign of the temperature contrast, $\Delta T$ indicates that the heat flux is higher when the probe is on the Al$_2$O$_3$ than on graphene, which is now acting like a sort of thermal barrier or, in other words, thermally resistive coating.

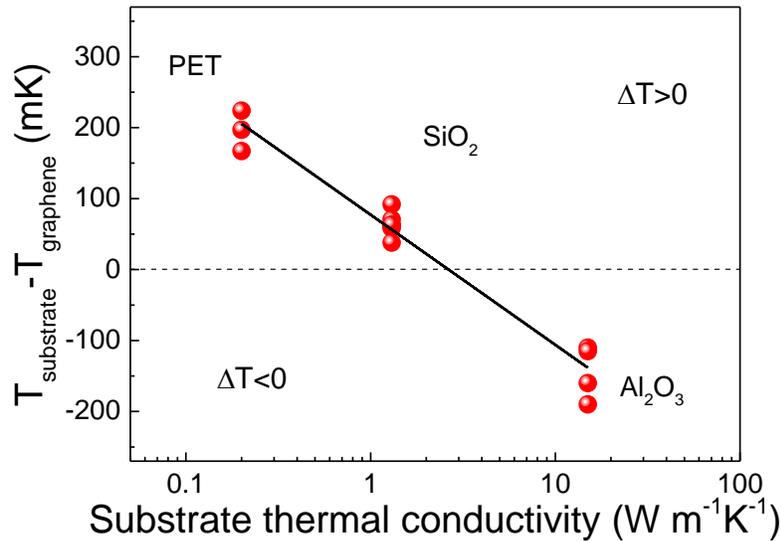

**Figure 6.** Summary of the temperature difference $T_{sub} - T_{GR} = \Delta T$ between the sensor temperature with the probe on the substrate and on one graphene layer, as a function of the thermal conductivity of the substrate. The black line is a log fit of the type $y = a \cdot \ln(x) + b$, where $a = -79.6 \pm 4.6\ mK$ and $b = 77.3 \pm 7.5\ mK$. The intercept at $y = 0$ is $x = 2.6 \pm 0.4 \frac{W}{mK}$.

The trend of $\Delta T$ as a function of the thermal conductivity of the substrate has been reproducibly observed by performing several measurements on different areas of each sample, as shown in Figure **6**.

This result indicates that the CVD graphene behaves as an ultrathin coating that improves heat dissipation on substrates whose thermal conductivity is equal or lower than that of SiO₂ ($k_{SiO2} = 1.4 \frac{W}{mK}$) while it behaves as a thin thermal barrier for more thermally conducting substrates. The line reported in Figure **6**, is a logarithmic fit of the type $y = a \cdot \ln(x) + b$ which intersects the $\Delta T = 0$ value at $k_{sub} = 2.6 \pm 0.4\ \frac{W}{mK}$. This is the simplest functional form that fits the data in this range and its physical meaning has to be investigated further. However, we do not expect it to have a wide range of validity, especially at higher conductivity values. With increasing values of the substrate thermal conductivity, the thermal spreading resistance of the system will decrease. Indeed, it has been shown [53] that for high values of the sample thermal conductivity the SThM tip is expected to progressively decrease its sensitivity. For example, in the case of a single isotropic sample, it will not be possible to distinguish thermal conductivity values above some tens of W/mK because the thermal resistance of the sample will be negligible with respect to that of the tip-sample contact (the two resistances are in series).

iii. *Thermal resistance maps and double-scan technique*

To make a more quantitative analysis, it is convenient to report the thermal maps in terms of the thermal resistance rather than of the temperature.

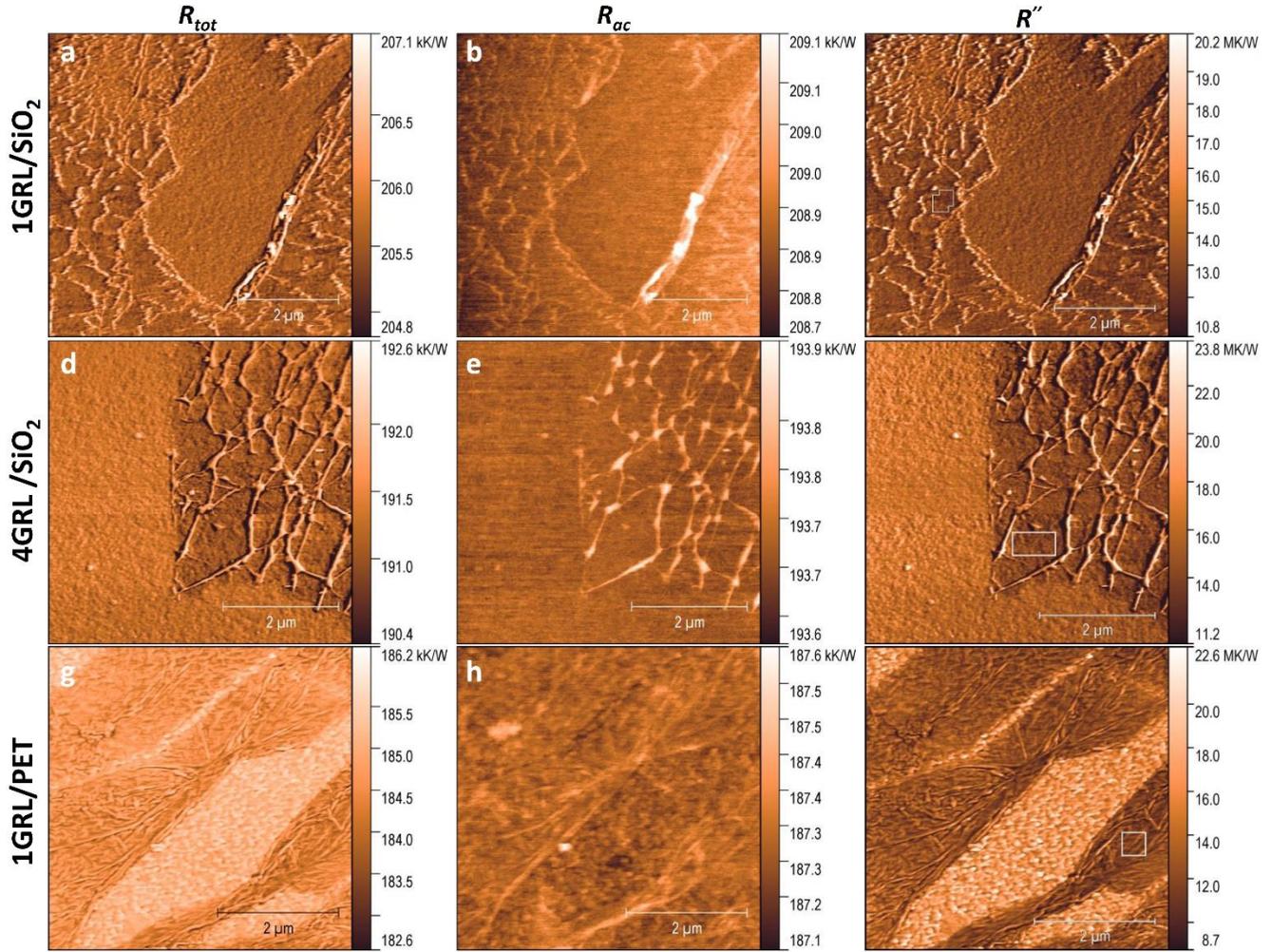

**Figure 7.** a: Total thermal resistance, $R_{tot}$ for the SThM probe in contact with 1GRL supported by SiO$_2$/Si substrate. b: Thermal resistance, $R_{a/c}$ obtained by a backward scan in the lift mode (tip close to the sample but not in contact), thus including only the heat dissipation through the air and the cantilever. c: Map of $R''$, obtained by the maps a and b and by applying to each pixel the formula: $\frac{1}{R_{tot}} = \frac{1}{R_{a/c}} + \frac{1}{R''}$, thus including only the tip-sample heat conduction. d, e, f and g, h, i: The same as in a, b, and c but for 4GRL supported by SiO$_2$/Si substrate and 1GRL supported by PET, respectively.

Since the resistance of the heater is known, we can calculate the heating power $\dot{Q}$ by using the Joule effect formula. From that, we can obtain the total thermal resistance of the system as $R_{tot} = \frac{T_H - T_0}{\dot{Q}}$. $R_{tot}$ can be expressed by the equation of the lumped elements circuit shown in Figure 4 b. We have seen that the circuit is represented by the parallel of two resistances: $R_{a/c}$ (that describes the contribution of heat conduction through the air and the cantilever) and the series $R_{t-s} + R_{GR} + R_{GR-Sub} + R_{Sp}^{Sub}$ that we can call, for simplicity, $R''$. $R''$ describes the heat conduction that occurs directly through the tip-sample channel and is present only when the probe is in contact with the sample. Thus, if the tip is very close to the sample but not touching it, the only contribution to the heat conduction will be, as a first approximation, given by $R_{a/c}$ only. In the light of this observation, we performed double scans by using the lift-mode technique. In the lift-mode scan, the forward trace is recorded with the tip in contact to the sample while the backward trace is obtained with the probe lifted to a certain height. This procedure is similar to that reported by Kim et al. [40] where, however, only line scans were performed instead of entire thermal maps as it is shown here. Different lift heights were explored, and we found that the optimal one is around 250 nm. Indeed, for lower lift heights the tip starts touching the sample during the backward scan due to the tip-sample electrostatic interaction, thus hindering the possibility of obtaining a clean map of $R_{a/c}$. On the other hand, for higher lift heights, $R_{a/c}$ is overestimated due to the excessive distance from the sample. From the height of 250 nm going down towards the contact, the tip-sample air transfer will still give a contribution, but it can be seen by performing retract measurements (see Supplementary Information file for more details) that this additional contribution is small compared to the total one. The retract measurements also confirmed that 250 nm is the minimum distance achievable from the experimental point of view to overcome electrostatic attraction of the probe to the sample. Figure **7** a reports the map of the total thermal resistance $R_{tot}$ for the 1GRL sample supported by SiO$_2$/Si. This map has been obtained by a forward scan, i.e. with the tip in contact with the sample. The graphene is visible mostly on the left and right side of the image, while the substrate corresponds to the flat central region. As expected, when scanning on the flat areas of the graphene, the probe features a lower thermal resistance than when it is on the SiO$_2$/Si substrate. Higher resistance values are obtained in correspondence of folds and impurities. Figure **7** b shows the map for $R_{a/c}$ obtained from the backward scan in the lift mode. This is the thermal signal that has been obtained when the probe is not in contact with the sample. The signal is obviously more blurred than before, but it is still possible to distinguish the most prominent topological features of the sample. This fact indicates that, as expected, the tip in this configuration is not only dissipating heat through the air and the cantilever, but that there is also an air-mediated heat transfer to the sample. This is exactly the contribution that we want to get rid of, in order to single out only the heat flux through the tip-sample contact. Then, since $\frac{1}{R_{tot}} = \frac{1}{R_{a/c}} + \frac{1}{R''}$, it is possible to determine $R''$, simply by inverting this formula. By applying the above formula using each pixel of the maps of Figure **7**a and b, we can calculate the map of $R''$, shown in panel c. It is possible to notice that, since $R_{tot}$ (panel a) and $R_{a/c}$ (panel b) have the same order of magnitude, $R''$ turns out to be about two orders of magnitude higher. This means that most of the heat generated at the heater is dissipated through the air and the cantilever. However, this does not hinder the capability of the probe to detect a clear temperature contrast when in contact with the sample. This fact is also confirmed by the much higher spatial resolution (a few tens of nm) that is achieved with the probe in contact than when it is lifted, as it can be seen by comparing panel a and b. On the other hand, it can also be shown that the spatial correlation of $R''$ with the topographic signal is not improved with respect to $R_{tot}$, but it is slightly lower (66.8% vs 68.2% in this case). This is due to the fact that the topological effects on the thermal maps will proportionally contribute more, as expected, to lower the correlation in the case of $R''$ than for $R_{tot}$, since these effects are, by definition, more relevant when the tip is in contact than when it is lifted. The value of $R''$ in correspondence of the masked graphene region is $(1.22 \pm 0.04) \times 10^7 \frac{K}{W}$ while it is $(1.28 \pm 0.03) \times 10^7 \ K/W$ on the substrate. Panels d, e and f report the maps of $R_{tot}$, $R_{a/c}$ and $R''$, respectively, for the 4GRL sample supported by SiO$_2$/Si. Considerations like those of the previous case hold here as well. Now $R''$ is $(1.41 \pm 0.08) \times 10^7 \frac{K}{W}$ when the probe scans in correspondence of the mask and $(1.65 \pm 0.09) \times 10^7 \frac{K}{W}$ when the probe is on the substrate. Panel g shows the $R_{tot}$ map for the 1GRL sample supported by PET. Darker areas with several wrinkles correspond to the graphene that is not continuous and features areas where the probe is in contact with PET (lighter areas). The $R_{a/c}$ map is reported in panel h. Edges, wrinkles and other topological irregularities are mostly visible. Panel i shows the $R''$ map where a clear contrast between graphene and PET can be noticed. $R''$ is $(1.18 \pm 0.10) \times 10^7 \frac{K}{W}$ on graphene (masked area) and $(1.69 \pm 0.18) \times 10^7 \frac{K}{W}$ on PET.

iv. *Analysis of the results for the monolayer supported by different substrates*

As in the case of the temperature variations, $\Delta T$ (reported in Figure **6**), also the thermal resistance decreases when passing from the substrate to the graphene in the case of the samples supported by PET and SiO₂ while it is higher on the graphene than on the substrate in the case of the Al₂O₃ substrate. This fact suggests that a convenient way to look at this type of systems is to regard the graphene deposited on the substrate as an effective material of thermal conductivity $k_{eff}$ ($k_{SiO2} < k_{eff} < k_{Al2O3}$) and thickness $t_{eff}$ ($t_{eff} > t_{graphene}$) in *perfect* contact with the substrate. The latter condition accounts for the graphene/substrate interface by increasing the thickness with respect to that of the bare graphene. $t_{eff}$ is therefore determined (similarly to what was done by Menges et al. [37]) as $t_{eff} = t_{graphene} + k_{eff} \cdot r_{eff}$, where $r_{eff}$ is an effective thermal boundary resistance parameter and $t_{graphene} = 0.34\ nm$. The quantity $k_{eff} \cdot r_{eff}$ is also known as the Kapitza length.

Since the thickness of the substrates is about 500 μm, the system in our case is equivalent to a layer of thermal conductivity $k_{eff}$ and thickness $t_{eff}$ in perfect contact to an infinite half-plane of thermal conductivity $k_{sub}$. The sum of the terms $R_{GR} + R_{GR-Sub} + R_{sp}^{Sub}$ can thus be described by the formula for the spreading resistance of a "compound half-plane" that, in the isoflux conditions, is [54]:

$$R_{sp}^q = \frac{\psi^q}{4k_{sub}a}$$

$$\tag{1}$$

Where $k_{sub}$ is the thermal conductivity of the substrate, $a$ the contact radius through which the heat is injected and

$$\psi^q = \frac{32}{3\pi^2}\left(\frac{k_{sub}}{k_{eff}}\right)^2 + \frac{8}{\pi}\left[1 - \left(\frac{k_{sub}}{k_{eff}}\right)^2\right] \cdot \int_0^\infty \frac{J_1^2(\zeta)d\zeta}{[1 + k_{eff}/k_{sub}\tanh(\zeta t_{eff}/a)]\zeta^2}$$

$$\tag{2}$$

where $J_1$ is the Bessel function of the first kind. $\psi_q$ is the dimensionless spreading resistance parameter that is defined as $\psi_q = 4k_{sub}aR_{sp}^q$ and its expression comes from that of the area-averaged temperature rise of the heat source area, $\bar{T}$ since the spreading resistance can be expressed by $R_{sp}^q = \frac{\bar{T}}{q\pi a^2}$, where q is the heat flux [54]. The isoflux condition has been chosen mainly for ease of calculation. However, it has been shown that the thermal spreading resistance in the isothermal conditions differs, at maximum, by 8 % [54]. Therefore

$$R'' = R_{t-s} + R_{sp}^q = \frac{r_{ts}}{\pi a^2} + R_{sp}^q$$

$$\tag{3}$$

where $r_{ts}$ is the interface resistance between the tip and the sample. Since in this model the heat "spreads" down into the sample through the contact area, it consequently accounts for the fact that heat transfer area between the graphene and the substrate is larger than the tip-sample contact radius, while the anisotropy of the graphene is embedded in the $k_{eff}$ and $t_{eff}$ parameters. To determine the characteristic parameters of the effective material, we make a couple of considerations: i) we assume that, in a single measurement, the contact area between the tip and the sample remains constant when passing from the graphene/substrate system to the bare substrate for that specific substrate. For example, the contact area for the tip on the graphene/SiO$_2$ system is the same as for the tip on the SiO$_2$ in the same measurement but it will be different for the case of the PET and Al$_2$O$_3$ substrates. This is reasonable because, as it can be seen from the topographic AFM images, the graphene, being thin and bendable, follows to a very good approximation the topography of the underlying substrate; ii) since the contact between the tip and the sample is formed by several nanocontacts, i.e. it is a multi-asperity contact [34], we assume that $r_{ts}$ is mainly determined by the morphology of the contact rather than the intrinsic properties of the two materials forming the contact. Therefore, it is kept constant when changing substrate. This is ascribed to the complex physical nature of the contact. Indeed, as shown in Gomes et al. [34], in the contact region the heat conduction occurs along several different channels: through mechanical contacts, water meniscus and ballistic conduction through the air.

The determination of the unique set of the three $k_{eff}$, $r_{eff}$ (or, equivalently, $t_{eff}$) and $r_{ts}$ values that reproduce the experimental results is a three-step process, that has been implemented by using a Matlab code.

Step 1. In principle, given a suitable couple of $k_{eff}$ and $r_{ts}$ values, by using the model for $R''$ reported in equation (3) and by spanning over a wide range of contact radii $a$, we can find for a given substrate the set of $r_{eff}$ values that give, separately, the correct experimental $R''_{gr/sub}$ and $R''_{sub}$ results. Then, we determine the only $r_{eff}$ value that matches the experimental data *with the same contact radius $a$* for both the tip/graphene/substrate and the tip/substrate contact. Indeed, we have two equations (one for $R''_{gr/sub}$ and one for $R''_{sub}$) and two unknown parameters ($a$ and $r_{eff}$).

Step 2. We determine many $r_{eff}$ values and, as a consequence, contact radius values, by spanning over a wide (with respect to all the possible realistic values) range of ($k_{eff}$, $r_{ts}$) values. For each different substrate, the result is a surface determined by all the sets of three ($k_{eff}$, $r_{eff}$, $r_{ts}$) parameters that match the experimental data for that given substrate. An example of these surfaces for the three substrates used in this work and for a specific set of experimental data is reported in Figure **8**.

Step 3. Then, we find the intersection between the three surfaces (one for each substrate) to determine the unique ($k_{eff}, r_{eff}, r_{ts}$) set. The black line in Figure **8** represents the intersection between the surfaces related to the SiO$_2$ and Al$_2$O$_3$ substrates, while the blue one is the intersection between those related to the SiO$_2$ and PET ones. The intersection between the two lines is the unique set of the ($k_{eff}, r_{eff}, r_{ts}$) values for that specific set of experimental data. Once the $r_{ts}$ value is determined, the three contact radii for each substrate are also consequently determined.

To summarize, we have six different measurements and six unknown parameters: $k_{eff}, r_{eff}, r_{ts}, a_{SiO2}, a_{PET}$ and $a_{Al2O3}$.

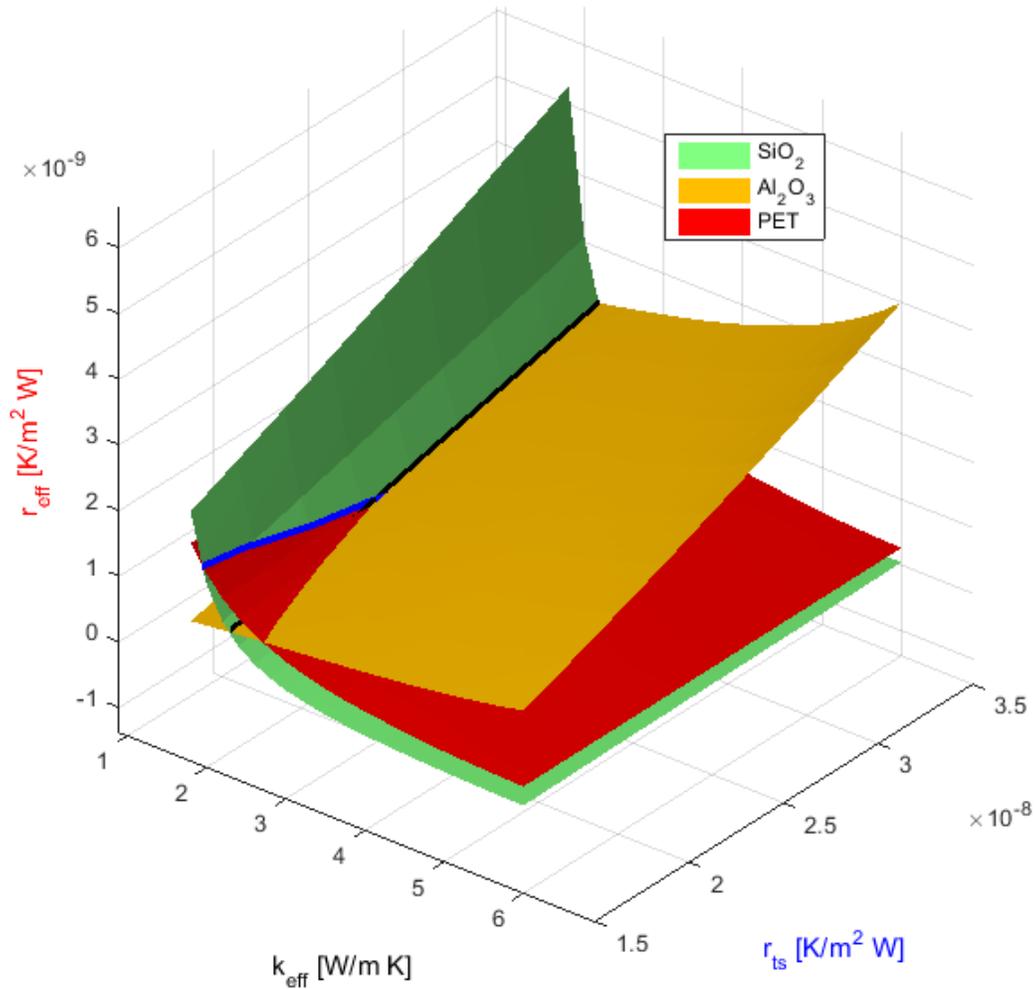

Figure 8. ($k_{eff}$, $r_{eff}$, $r_{ts}$) surfaces representing the solutions, for each substrate and for a particular couple of experimental values of $R''_{gr/sub}$ and $R''_{sub}$, of equation (3) by imposing the same contact radius $a$. The experimental values used in these calculations are: $R''_{gr/PET} = 1.25 \times 10^{-7} \, K/W$, $R''_{PET} = 1.67 \times 10^{-7} \, K/W$, $R''_{gr/SiO2} = 1.22 \times 10^{-7} \, K/W$, $R''_{SiO2} = 1.28 \times 10^{-7} \, K/W$, $R''_{gr/Al2O3} = 4.25 \times 10^{-6} \, K/W$ and $R''_{Al2O3} = 4.02 \times 10^{-6} \, K/W$.

We performed the above procedure by using different sets of data corresponding to different masked areas on the samples and we obtained $k_{eff} = 2.5 \pm 0.3 \, W/mK$, $t_{eff} = 3.5 \pm 0.3 \, nm$ and $r_{ts} = (2.4 \pm 0.6) \times 10^{-8} \, Km^2/W$. The corresponding $r_{eff}$ values are $r_{eff} = (1.2 \pm 0.2) \times 10^{-9} Km^2/W$. The contact radii were found to be in the 30-50 nm range for the SiO$_2$ and Al$_2$O$_3$ substrates, while higher (about 80 nm) for PET.

As expected, the $k_{eff}$ values are between the thermal conductivity of SiO$_2$ and Al$_2$O$_3$, but closer to that of SiO$_2$ and the obtained value perfectly coincides, within the uncertainty bar, with that of the intersection between the fit line in Figure 6 and $\Delta T = 0$, that was found to be $x = 2.6 \pm 0.4 \, W/mK$, indicating that the fitting procedure could be a good method for a quick estimation of $k_{eff}$. Moreover, it is worth pointing out here that this value is related to the heat injection *perpendicular* to the plane. Therefore, it should not be compared to the in-plane one for the supported graphene which can even be of the order of a few hundreds of $W/mK$ [19, 7]. As for the $r_{eff}$ value which determines the effective thickness, $t_{eff}$ of the graphene coating, it has the physical dimensions of a thermal boundary resistance. A comparison between this value and those reported for the thermal boundary resistance between graphene and different substrates [55,56] has some limitations because in our model the graphene and the interface form a single entity (indeed it would be problematic to define the *c*-axis thermal conductivity for a single graphene layer). Nevertheless, we can notice that the order of magnitude of $r_{eff}$ is in the realistic range for the thermal boundary resistances [57] and that the obtained value is very close to the range reported by ref. [55] for a graphene/SiO$_2$ interface, but lower than others [56,58,59]. Values similar to ours have also been reported for the carbon nanotube (CNT)/SiO$_2$ interface [60] and for the graphene/oil interface [61]. The thermal boundary resistance values for other carbon compounds like diamond [63], metallic single-wall CNTs [64] and graphite [10,65] are close to the upper bound of thermal resistances found for graphene, i.e. of the order of 10$^{-8}$ Km$^2$/W. It is also worth recalling here that the fact that $r_{eff}$ and, consequently, $t_{eff}$ is assumed to be constant on different substrates is the most severe assumption. However, we believe it is sensible in this case because, as stated in the beginning, the presence of a graphene-substrate adsorbate layer [19,20,21] caused by the wet conditions for the sample preparation will tend to make the interface properties similar among different substrates. Finally, we checked in particular that the contact radius for the Al$_2$O$_3$ case (that was found to be about 40 nm) is larger than the phonon mean free path, because the expression of eq. 1 is based on the diffusive heat conduction. We estimated the phonon mean free path, $l_{ph}$ from the formula $\Lambda = \frac{1}{3} C \rho v l_{ph}$, where $\Lambda$ is the thermal conductivity, $C$ is the specific heat, $\rho$ is the density, $v$ the sound velocity. The material properties were taken from the literature [62]. We obtained $l_{ph} \cong 3.3\ nm$, much smaller than the obtained tip-sample contact radius. Even though the kinetic expression used here for the calculations might underestimate the mean free path by a factor of 4-5, the diffusive heat conduction conditions would be met anyway.

v.   *Analysis of the results for 2 and 4 layers supported by SiO$_2$.*

The two-layer and four-layer samples have been obtained by multiple transfer procedures of single CVD layers, i.e. each layer has been subsequently stacked one on top of the other. Therefore, their properties are expected to be quite different from those of the exfoliated bi- and four-layer graphene. In our model of graphene as a thermal coating in perfect contact with the substrate, the addition of one layer can be regarded as equivalent to the addition of one layer of the effective material with thermal conductivity $k_{eff}$. The only difference is that now, besides $r_{eff}$, there is an additional interface parameter that describes also the interaction between different graphene layers and that we name $r_{mlg-eff}$. Therefore, the effective thickness of each additional layer after the first will in principle be different from that of the first one. The total effective thickness can thus be expressed as $t_{n-eff} = n \cdot t + k_{eff} \cdot [r_{eff} + (n-1) \cdot r_{mlg-eff}]$ where $n$ is the number of stacked graphene layers. By using the $k_{eff}$ values found for the monolayer case, we obtained $t_{2-eff} = 7.6 \pm 3.5\ nm$ (corresponding to $r_{mlg-eff} = (1.6 \pm 1.5) \times 10^{-9}\ K \cdot m^2/W$) and $t_{4-eff} = 26 \pm 12\ nm$ (corresponding to $r_{mlg-eff} = (3.0 \pm 1.8) \times 10^{-9}\ K \cdot m^2/W$). The results are reported in Figure **9.** The error bars are rather large because these results have been obtained by averaging over many measurements obtained in different regions of the samples and therefore are affected by local inhomogeneities. By looking at the effective thickness per number of layers, $t_{n-eff}/n$ (see inset to Figure **9**), it is possible to notice that the stacking of the second graphene layer only slightly improves the heat conduction because $t_{n-eff}/n$ for two layers ($t_{2-eff}/2 = 3.8 \pm 1.7\ nm$) is very close to that of a single layer ($t_{eff} = 3.5 \pm 0.3\ nm$). On the other hand, when 4 layers are stacked, a noticeable improvement of the heat conduction can be noticed. In this case, 4 graphene layers are equivalent to about 7.4 effective material layers and $t_{4-eff}/4 = 6.5 \pm 3\ nm$. It might seem counterintuitive that the heat dissipation improves when the effective thickness of the conductive coating increases, but it is worth recalling here that, since the substrate ($SiO_2$) is less conducting than the coating material, an increase of the effective thickness of the conductive coating will decrease the total spreading resistance of the compound half-plane [54]. Furthermore, let us note that even though the graphene/graphene interface is expected to be more efficient than the graphene/substrate one [10,27], this improvement looks still rather weak in the case of 2 layers, where the interface between the second and first layer is most probably still influenced by the substrate. Then, when the number of layers increases to 4, the improvement is clear. Of course, an exfoliated bi- or four-layer sample is expected to dissipate much more, not only because of the intrinsic higher quality of the individual layers, but also because of the better thermal interface between the different graphene planes due to the non-random stacking and to the absence of adsorbates between the planes.

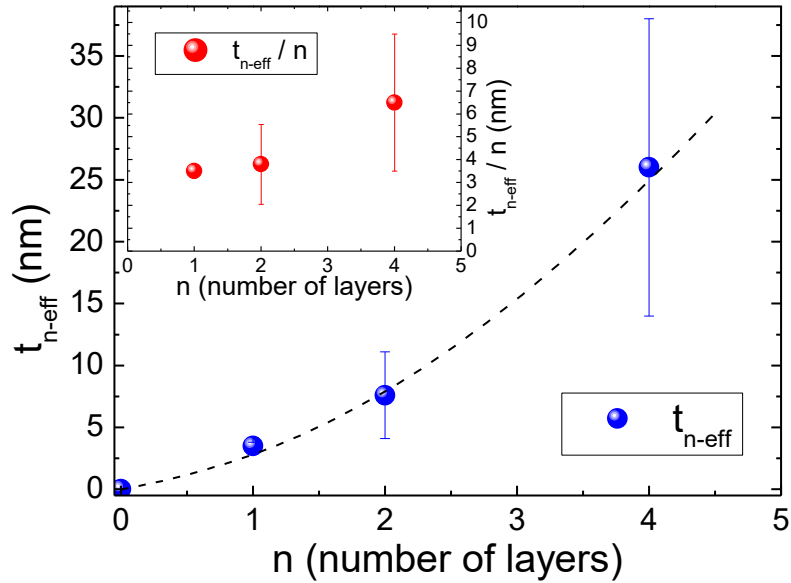

**Figure 9.** Main panel: Effective thickness of 1GRL, 2GRL and 4GRL supported by SiO$_2$/Si as a function of the number of layers. Inset: The same as in the main panel but now the effective thickness is normalized to the number of layers.

The increase of the effective thickness of the coating material in the case of the 2GRL and 4GRL samples is a possible way to model the decrease of the spreading resistance with increasing number of layers. Alternatively, 2GRL and 4GRL could of course be considered as materials with a different effective thermal conductivity, $k_{2-eff}$ and $k_{4-eff}$ ($>k_{eff}$), and their relevant effective thickness. For precisely obtaining these values, we should perform SThM measurements on these samples, supported by at least two other different substrates, like PET and Al$_2$O$_3$, but this is beyond the scope of the present work and is the subject of future analyses. At present, we can however safely determine a lower bound for $k_{2-eff}$ and $k_{4-eff}$, by using the experimental data for 2GRL and 4GRL supported by SiO$_2$ and by conservatively supposing that $\Delta T$ for the 2GRL and 4GRL supported by $Al_2O_3$ stays unchanged with respect to the 1GRL/$Al_2O_3$ sample, i.e. $\Delta T_{2GRL/Al2O3} = \Delta T_{4GRL/Al2O3} = \Delta T_{1GRL/Al2O3}$. By connecting these values, we obtain the dashed blue and olive lines reported in Figure **10**, respectively. The intercept is $3.3\ W/mK$ (dashed blue line in Figure **10**) for 2GRL and $5.7\ W/mK$ for 4GRL (dashed olive line). In other words, by performing the same procedure shown for the 1GRL sample on the 2GRL and 4GRL ones, we would expect to obtain *at least* $k_{2-eff} \cong 3.3\ W/mK$ and $k_{4-eff} \cong 5.7\ W/mK$. The data are reported as symbols in the inset to Figure **10**. It is interesting to compare this result with the work of Jang et al. [66], where the thermal properties of graphene encased in $SiO_2$ have been studied for different number of layers. Contrary to what observed for suspended graphene [10], and like the results shown here, an increase of the thermal conductivity has been measured with increasing number of layers. There, the effect was ascribed to the presence of the oxide (on both sides of the sample) that suppresses that thermal conduction over a characteristic length. A quantitative comparison of the obtained thermal conductivity values is not possible because in that case the graphene was exfoliated, and the in-plane conduction was probed while we are here sensitive to an overall effective conductivity. However, a similar effect is very likely to occur here as well. The best fit of the data is obtained with a 2[nd] order polynomial fit (dashed red line). At about 10 stacked layers the conductivity turns out to be $k_{eff} \cong 20\ W/mK$. However, since $k_{eff}$ is expected to saturate with increasing number of layers, we also tried to fit our data with the model reported in equation (2) or ref. [66], in order to better estimate the expected trend of the data. In this model, we have three free parameters: the thermal conductivity for thin flakes, $k_0$, the "bulk" thermal conductivity, $k_{bulk}$, and the characteristic penetration of the detrimental effects of the substrate, $\delta$. First, we impose, of course, $k_0 = 2.5\ W/mK$. Then, since we observed experimentally that the conduction properties for the 4-layer sample are better than the 2-layer one, we conservatively limit the upper bound for $\delta$ to 3 layers. In this case we get that the thermal conductivity at 10 layers is about 15 W/mK and the "bulk value" is 30 W/mK. These values would be higher with a larger $\delta$. For example, if we allowed $\delta = 4$, we would get a thermal conductivity of 17 W/mK for 10 layers with a bulk value of almost 50 W/mK. Even though we don't have enough experimental information on this characteristic length, and the samples here are different from those of ref. [66], it should be kept in mind that the estimated $k_{2-eff}$ and $k_{4-eff}$ values have been obtained in the most conservative way and represent a lower bound for $k_{eff}$.

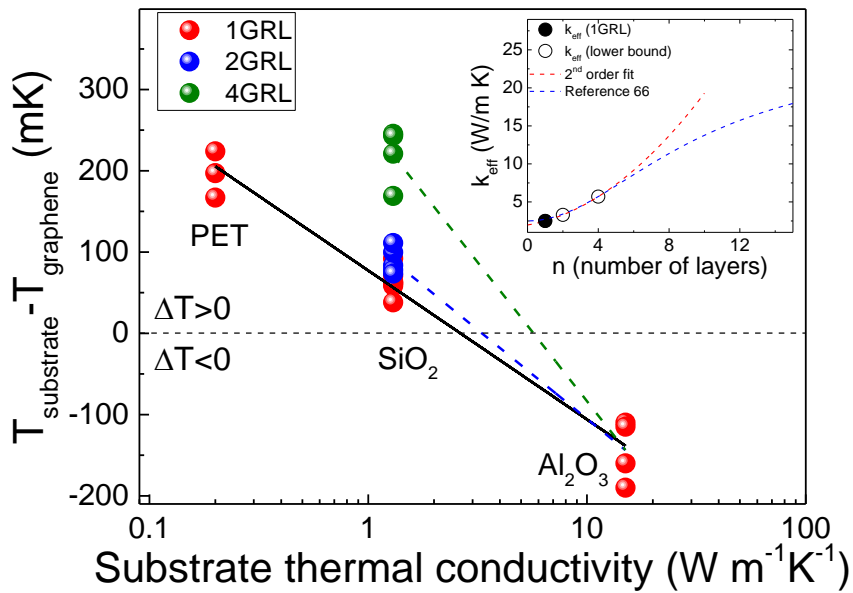

**Figure 10.** Main panel: Summary of the temperature difference $T_{sub} - T_{GR} = \Delta T$ between the sensor temperature with the probe on the substrate and on one (red), two (blue) and four (olive symbols) graphene layers, as a function of the thermal conductivity of the substrate. The black line is a log fit of the type $y = a \cdot \ln(x) + b$, where $a = -79.6 \pm 4.6 \, mK$ and $b = 77.3 \pm 7.5 \, mK$. The intercept at $y = 0 \, mK$ is $x = 2.6 \pm 0.4 \, W/m \cdot K$. Blue and olive dashed lines are analogous log fits connecting the average value of the 2GRL and 4GRL samples supported by SiO$_2$, respectively to that of 1GRL supported by Al$_2$O$_3$ (see text for details). Inset: Calculated (full symbol) and estimated lower bound (open symbols) values of $k_{eff}$ as a function of number of stacked graphene layers The red line is a 2$^{nd}$-order polynomial fit to the data while the blue one is a fit performed by using equation (2) of ref [66].

Thus, with a thickness of a few nanometers, the effective thermal conductivity is expected to reach some tens of $W/m \cdot K$, which is comparable to that of thin films of metals like Al, Cu and Au, where the thermal conductivity can be 20% of the bulk value when the thickness is of the order of 100 nm [27]. This is the case, for example, of gold thin films deposited on etched Si with a thickness comparable to the electronic mean free path [67,68], while for Al thin films, we would expect $k \cong 26 - 48\ W/m \cdot K$ [27]. Finally, it is important to recall that these multilayer systems are highly anisotropic, with an in-plane conductivity that can be orders of magnitude higher than the out-of-plane one. Therefore, they can be very useful for achieving a high in-plane heat dissipation with a very small thickness of coating material.

**CONCLUSIONS**

In conclusion, we have reported on the first SThM results on CVD graphene supported by different substrates (SiO₂, PET, Al₂O₃). For the SiO₂ substrate, 2- and 4-layer samples were investigated as well. The SThM measurements were performed with a double-scan technique to get rid of the heat dissipation through the air and the cantilever. Then, by using a simple lumped-elements model for the probe/sample system, along with the expressions of the spreading resistance in a compound half plane, we developed a multi-step analysis that allows determining the effective thermal conductivity (and effective thickness) of thermally conductive coatings of nanometric thickness. In the specific study reported here, we have shown that the single CVD graphene layer behaves, for heat injection perpendicular to the graphene planes, as a thermal coating equivalent to an effective material of conductivity $k_{eff} = 2.5 \pm 0.3\ W/mK$ and thickness $t_{eff} = 3.5 \pm 0.3\ nm$ in *perfect* contact with the substrate. It is thus conductive in the case of SiO₂ and PET substrates ($k_{eff} > k_{sub}$) while it is resistive in the case of Al₂O₃ ($k_{eff} < k_{sub}$). We have also shown that the heat conduction properties improve with increasing number of layers on SiO₂ and that, with a technologically achievable number of layers, the effective thermal conductivity is expected to be comparable to that of some thin films of metals with a thickness one order of magnitude higher, thus confirming the interest for the application of the industrially viable CVD graphene sheets. This improvement is due to both the fact that with increasing number of layers the detrimental effect of the substrate decreases and that a thicker thermal coating deposited on a resistive substrate will reduce the total thermal spreading resistance. This new method is very helpful for determining the equivalent thermal coating properties of 2D materials and can be used for the design of applications for thermal management and heat dissipation in nanoelectronics devices and thermally conductive coatings. These results also show the importance of carefully determining and investigating the properties of graphene and graphene-related in the specific situations in which they are employed.

Corresponding author

*e-mail: mauro.tortello@polito.it

Acknowledgements

This work has received funding from the European Research Council (ERC) under the European Union's Horizon 2020 research and innovation program, grant agreement 639495 — INTHERM — ERC-2014-STG and the European Union's Horizon 2020 Framework Program under Grant Graphene Core2 n°785219 Graphene Flagship.

Author contributions

M.T. and A.F. designed the research. I.P., K.Z.-C. and W.S. prepared and characterized the samples. M.T. performed the SThM measurements and elaborated the data. M.T. and R.S.G. developed the multi-step analysis method. M.T. mainly wrote the paper. All authors participated to the discussion of the results.

52. Kretinin, A. V.; Cao, Y.; Tu, J. S.; Yu, G. L.; Jalil, R.; Novoselov, K. S.; Haigh, S. J.; Gholinia, A.; Mishchenko, A.; Lozada, M.; Georgiou, T.; Woods, C. R.; Withers, F.; Blake, P.; Eda, G.; Wirsig, A.; Hucho, C.; Watanabe, K.; Taniguchi, T.; Geim, A. K.; Gorbachev R. V. Electronic Properties of Graphene Encapsulated with Different Two-Dimensional Atomic Crystals. *Nano Lett.* **2014**, *14*, 3270−3276.

53. Tovee P.; Pumarol, M.; Zeze, D.; Kjoller, K.; Kolosov O. Nanoscale Spatial Resolution Probes for Scanning Thermal Microscopy of Solid State Materials, *J. Appl. Phys.* **2012,** *112,* 114317.

54. Yovanovich, M. M.; Culham, J. R.; Teertstra, P. Analytical Modeling of Spreading Resistance in Flux Tubes, Half Spaces, and Compound Disks *IEEE transactions on components, packaging, and manufacturing technology* **1998,** *21,* 168-176.

55. Chen, Z.; Jang, W.; Bao, W.; Lau, C. N.; Dames, C. Thermal Contact Resistance Between Graphene and Silicon Dioxide. Appl. Phys. Lett. **2009**, *95,* 161910.

56. Schmidt, A. J.; Chen, X.; Chen, G. Pulse Accumulation, Radial Heat Conduction, and Anisotropic Thermal Conductivity in Pump-Probe Transient Thermoreflectance. *Rev. Sci. Instrum.* **2008,** *79,* 114902.

57. Wang, H.; Xu, Y.; Shimono, M.; Tanaka, Y.; Yamazaki, M.; Computation of Interfacial Thermal Resistance by Phonon Diffuse Mismatch Model. *Materials Transactions* **2007,** *48,* 2349-2352.

58. Mak, K. F.; Lui, C. H.; Heinz, T. F. Measurement of the Thermal Conductance of the Graphene/SiO$_2$ Interface. *Appl. Phys. Lett.* **2010,** *97,* 221904.

# Supporting Information:

## Chemical Vapor Deposition Graphene as a Thermally Conducting Coating

Mauro Tortello*,[1], Iwona Pasternak[2], Klaudia Zeranska-Chudek[2], Wlodek Strupinski[2], Renato S. Gonnelli[1], Alberto Fina[1]

1. Dipartimento di Scienza Applicata e Tecnologia, Politecnico di Torino, 10129 Torino, Italy

2. Faculty of Physics, Warsaw University of Technology, Koszykowa 75, 00-662 Warsaw Poland

*To whom correspondence should be addressed: mauro.tortello@polito.it

# Retract measurements

The variation in the thermal signal due to the tip-sample heat transfer from the contact-mode operating condition (deflection signal = 0.5 V) to a retraction of 250 nm (corresponding to the lift mode), is small compared to the total one. At about 4 μm distance from the sample (maximum distance experimentally achievable for the retraction of the probe), the heat transfer to the sample is still far from being negligible. This can be seen by the fact that the thermal signal at -4 μm is still varying, even though the magnitude of the slope of *($V_s$-$V_r$) vs z* is decreasing with increasing tip-sample distance.

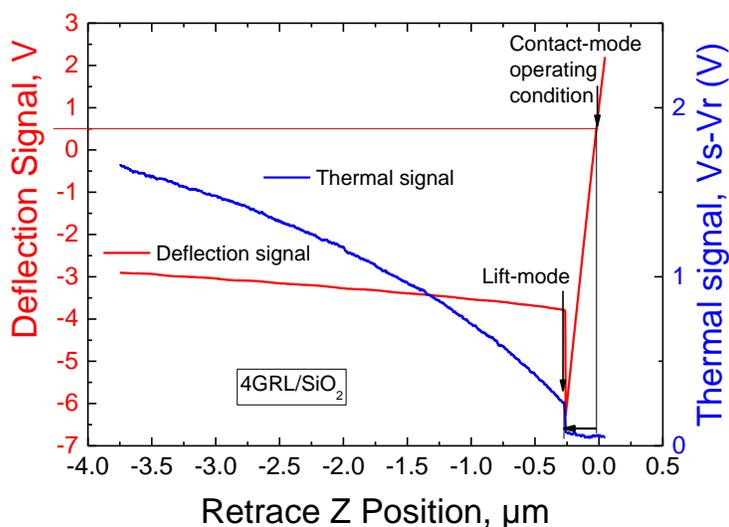

Figure **S1.** Retract measurement on a 4-layer sample supported by SiO$_2$. Red: deflection signal. *0.5V* corresponds to the force set for the SThM maps acquisition. Blue: SThM thermal signal.

Retract measurements also show that 250 nm is the minimum achievable distance from the experimental point of view. Two examples are reported below. When scanning in the lift mode with a 250 nm distance, the tip-sample height is a few percent within the minimum achievable distance below which electrostatic effects keep attracting the tip to the sample.

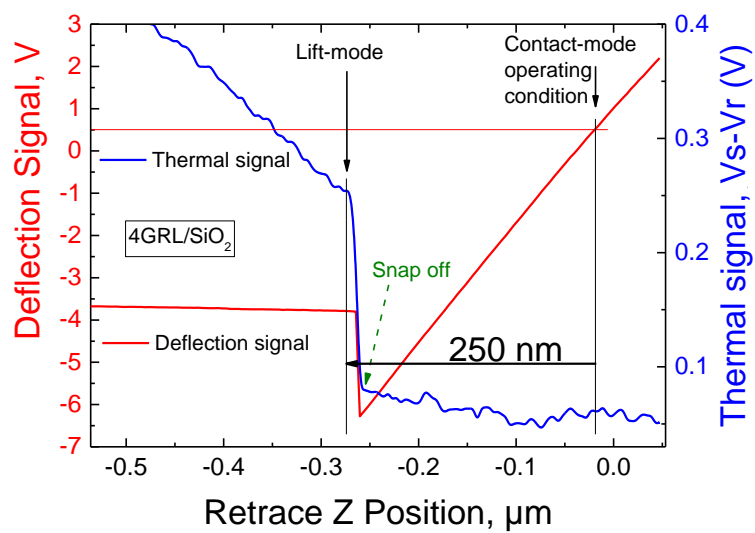

Figure **S2**. The same as Figure **S1** but magnified. Note the thermal signal corresponding to the deflection setpoint of 0.5 V, at about *z=0 nm* and that in the lift mode, at about *z=-250 nm*.

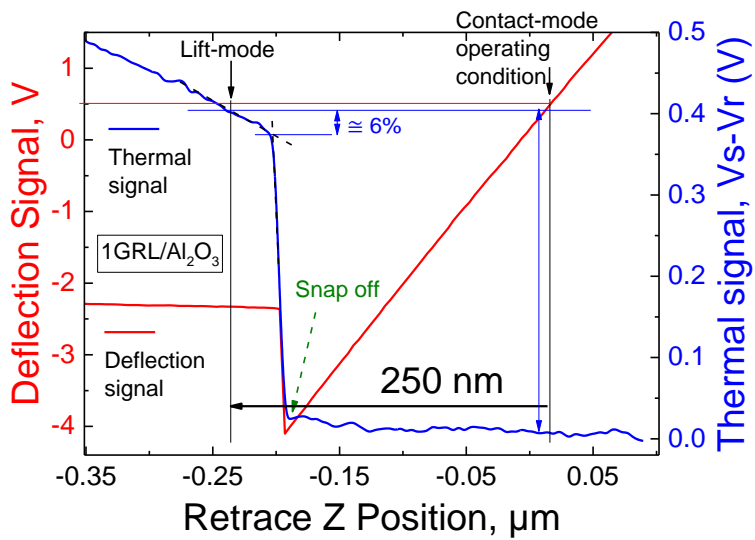

Figure **S3**. Retract measurement on a 1-layer sample supported by $Al_2O_3$. Red: deflection signal. *0.5V* corresponds to the force set for the SThM maps acquisition. Blue: SThM thermal signal. Note the thermal signal corresponding to the deflection setpoint of 0.5 V, at about *z=0 nm* and that in the lift mode, at about *z=-250 nm*.